\documentclass[preprint2]{aastex6}

\usepackage{amssymb,amsmath,graphicx,natbib,hyperref}
\usepackage{booktabs}
\usepackage{url}
\usepackage{graphicx}
\usepackage{color}
\usepackage{txfonts}

\newcommand{\Ha}{$\rm{H}\alpha$}
\newcommand{\Hb}{$\rm{H}\beta$}

\shorttitle{MZR does not depend on SFR}

\begin{document}

\title{Separate ways: The Mass-Metallicity Relation does not strongly correlate with Star Formation Rate in SDSS-IV MaNGA galaxies} 

\email{jbarrer3@jhu.edu}

\author{J.K. Barrera-Ballesteros\altaffilmark{1}, S.F. S\'anchez\altaffilmark{2}, T. Heckman\altaffilmark{1}, G. A. Blanc\altaffilmark{3,4} \& the MaNGA team}
\altaffiliation{Department of Physics \& Astronomy, Johns Hopkins University, Bloomberg Center, 3400 N. Charles St., Baltimore, MD 21218, USA}
\altaffiliation{Instituto de Astronom\'ia, Universidad Nacional Aut\'onoma de M\'exico, A.P. 70-264, 04510 M\'exico, D.F., M\'exico}
\altaffiliation{Observatories of the Carnegie Institution for Science, 813 Santa Barbara St, Pasadena, CA, 91101, USA}
\altaffiliation{Departamento de Astronom\'ia, Universidad de Chile, Camino del Observatorio 1515, Las Condes, Santiago, Chile}

\begin{abstract}
We present the integrated stellar mass --metallicity relation (MZR) for more than 1700 galaxies included in the integral field area SDSS-IV MaNGA survey. The spatially resolved data allow us to determine the metallicity at the same physical scale (effective radius, $\mathrm{R_{eff}}$ in arcsecs) using a heterogeneous set of ten abundance calibrators. Besides scale factors, the shape of the MZR is similar for all calibrators, consistent with those reported previously using single-fiber and integral field spectroscopy. We compare the residuals of this relation against the star formation rate (SFR) and specific SFR (sSFR). We do not find a strong secondary relation of the MZR with either SFR or the sSFR for any of the calibrators, in contrast with previous single-fiber spectroscopic studies. Our results agree with an scenario in which metal enrichment happens at local scales, with global outflows playing a secondary role in shaping the chemistry of galaxies and cold-gas inflows regulating the stellar formation. 
\end{abstract}
\section{Introduction}

Current chemical content in nearby galaxies is the consequence of the star-formation and chemical enrichment history. In particular, the observed oxygen abundance is consequence of cosmological evolution. As a result, these abundances present strong correlations with other parameters such as the total stellar mass.  

Although the relation between the galaxy luminosity and metallicity has been know for decades \citep[e.g.,][]{1992MNRAS.259..121V}, the mass--metallicity relation (MZR) was introduced by \cite{2004ApJ...613..898T}. It exhibits a tight correlation between the integrated stellar mass and the average oxygen abundance of galaxies: as stellar mass increases, the metallicity increases reaching a saturation at high stellar masses. They derived the MZR with a tight dispersion ($\sim$0.1 dex) for $\sim$40,000 galaxies extracted from the SDSS spectroscopic sample at z$\sim$0.1. Although its functional form seems to depend on the adopted abundance calibrator \citep[e.g.,][]{2008ApJ...681.1183K}, it is rather stable when using single aperture spectroscopic data or spatial resolved information \citep[e.g.,][]{2012ApJ...756L..31R, 2014A&A...563A..49S}. 

The MZR was interpreted by \cite{2004ApJ...613..898T} as the result of galactic outflows regulating the metal content of the inter-stellar medium. Alternatively, \cite{2012ApJ...756L..31R} show that the integrated relation is easily derived from a new, more fundamental, relation between the stellar mass density and the local oxygen abundance. This relation has been confirmed by \cite{2013A&A...554A..58S} and recently using MaNGA data by \cite{2016MNRAS.463.2513B}. In this scenario the stellar mass growth and the metal enrichment are both dominated by local processes, in-situ star formation, with a little influence of outflows or radial migrations.

Different authors have investigate a possible dependence of the MZR with the SFR \citep[e.g.,][]{2008ApJ...672L.107E,2010MNRAS.408.2115M, 2010A&A...521L..53L,2016ApJ...827...35T}. In different degrees, these studies show that at a fixed stellar mass, galaxies with stronger SFR exhibit lower oxygen abundances. Although the adopted functional form for this secondary relation is different depending on the  study, the conclusions are similar. Since oxygen abundance is enhanced due to star-formation, which in turn is directly related to the production of type-II SN, the proposed secondary correlation is therefore not quite intuitive. These studies are based on sub samples of the same observational dataset, the SDSS spectroscopic survey at z$\sim$0.1. Despite the fact of the application of aperture corrections \citep{2004MNRAS.351.1151B}, the spectroscopic information is affected by strong aperture effects \citep[e.g.,][]{2016ApJ...826...71I,2013A&A...553L...7I,2016A&A...586A..22G}.

\cite{2013A&A...554A..58S} could not confirm this secondary relation using integral field spectroscopic data covering the full optical extension of the galaxies, extracted from the CALIFA dataset \citep{2012A&A...538A...8S}. This result was confirmed recently with more statistics by \cite{2015Galax...3..164S}. In a recent study using drift-scan integrated spectra, \cite{2013A&A...550A.115H} show that a secondary relation of the MZR with the SFR is not present. Indeed, \cite{2012ApJ...756L..31R} had already shown that the relation with the sSFR (using as a proxy the \Ha\, equivalent width [EW(\Ha)]) of the local MZ relation does not present a secondary trend, but follows the primary relation between the SFR and stellar mass, as studied in detail by \cite{2013A&A...554A..58S} and more recently by \cite{2016arXiv160202770C}. These results were also discussed in \cite{2014ApJ...797..126S}, who divided the data presented by \cite{2013A&A...554A..58S} in mass bins and found a correlation between the metallicity and the sSFR in each of those bins. Those correlations are easily explained as a consequence of the combination between the SFR-mass and mass-metallicity relations. These secondary dependences disappear if the primary dependence of the mass with metallicity is removed. Finally, \cite{2012ApJ...745...66M} showed that this secondary relation is not shown in their data, rather they propose a secondary relation with the gas fraction. Thus, the secondary relation of the SFR and the MZR has been observed, so far, only in aperture-based spectroscopic observations.

In this article we explore the MZR and its possible dependence with the SFR and sSFR for more than 1700 galaxies included in the spatially-resolved spectroscopic MaNGA survey \citep{2015ApJ...798....7B}. The distribution of the article is as follows: in Section \ref{sec:sample} we present our sample of galaxies, an overview of the dataset as well as a brief description of the ten calibrators used to derive the oxygen abundance; the statistical wealth of the data allow us to present in Section \ref{sec:MZ} the MZR at a fixed physical scale (i.e., $\mathrm{R_{eff}}$); we explore the possible dependence of its residuals with the SFR and sSFR in Section \ref{sec:residuals} as well as the impact of the stellar mass and aperture effects in these residuals; these results are discussed in Section \ref{sec:Dis}; finally we present the main conclusions of this work in Section \ref{sec:con}.

\section{Sample and data}
\label{sec:sample}

\subsection{The MaNGA sample}
\label{sec:Manga}
%
\begin{figure}
\includegraphics[width=\columnwidth]{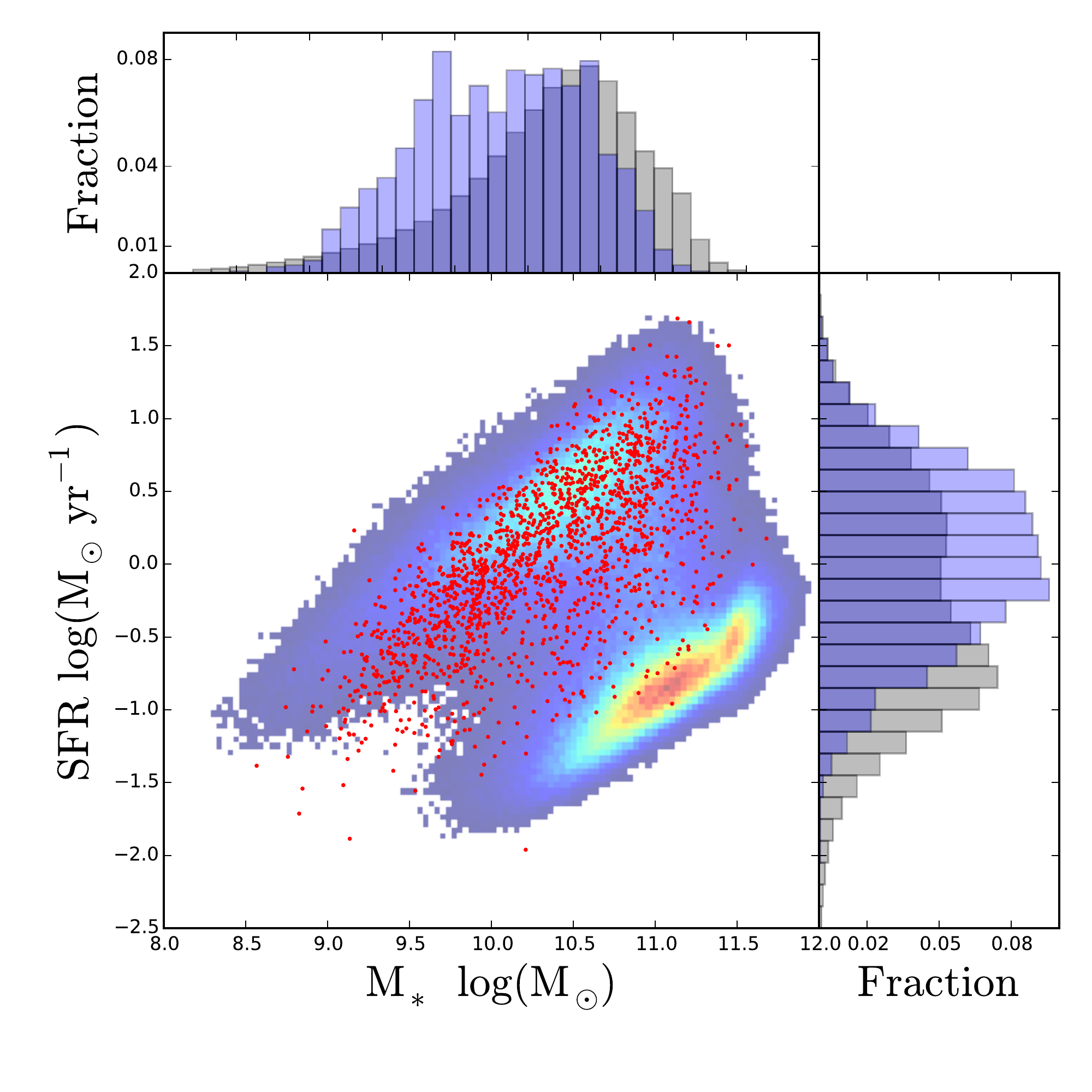}
\caption{Comparison of our sample of galaxies and the SDSS-DR7 sample in the stellar mass - SFR plane. The background distribution shows the well-know star-formation and retired sequence in the SDSS galaxies. Red points show the total SFR and stellar masses for our selected sample (1704 galaxies). On the horizontal and vertical histograms we compare the stellar mass and SFR distributions for these two samples. In these histograms, grey and blue panels represent the SDSS and our sample, respectively.} 
\label{fig:Sample}
\end{figure}
For this study we use the sample observed by the MaNGA survey until June 2016 (2730 galaxies at redshift 0.03$<z<$0.17). The goal of the ongoing MaNGA survey is to observe approximately 10.000 targets, detailed description of the selection parameters can be found in \cite{2015ApJ...798....7B}. A detailed a description of the sample properties is found in \cite{2016AAS...22733401W}. The MaNGA survey is taking place at the 2.5 meter Apache Point Observatory \citep{2006AJ....131.2332G}. Observations are carried out using a set of 17 different fiber-bundles science IFU \citep{2015AJ....149...77D}. These IFS feed two dual channel spectrographs \citep{2013AJ....146...32S}. Details of the survey spectrophotometric calibrations can be found in \citep{2016AJ....151....8Y}.Observed datacubes are reduced by a dedicated pipeline described in \cite{2016AJ....152...83L}. This sample covers a wide range of parameters (e.g, stellar mass, SFR and morphology), providing a unique view of galactic properties in the Local Universe.

To extract the two-dimensional physical properties from the reduced datacubes we used the analysis pipeline PIPE3D \citep[][]{2016arXiv160201830S}. For a detailed description on the fitting procedure and uncertainties determination see \cite{2015arXiv150908552S}. An overview on how this pipeline extracts the maps of the physical properties from ionized-gas emission lines datacubes is described in \cite{2016MNRAS.463.2513B}. Prior to deriving the characteristic oxygen abundances in each galaxy, we first select those regions (i.e., spaxels) that meet the following criteria:\textit{(a)} flux ratios ([OIII]/\Hb\, and [NII]/\Ha) lying in the star-forming region of a BPT \citep{1981PASP...93....5B} diagnostic diagram (i.e., below the \cite{2003MNRAS.346.1055K} demarcation line) and \textit{(b)} an EW(\Ha) larger than 6\,\AA.  These combined criteria ensure that the ionization is due to young stars \citep[e.g.,][]{2011MNRAS.413.1687C, 2014A&A...563A..49S}. We convert the luminosity of the \Ha\, emission line for each of the spaxels to their SFR using the relation presented in \cite{1998ApJ...498..541K}. Then, we co-add together those spaxels with ionization compatible with star-formation to derive the integrated SFR. The result of obtaining the total SFR using this procedure does not change substantially compared to (for instance) the total SFR from the integrated H$\alpha$ flux within the IFU's field of view \citep[e.g.,][]{2015A&A...584A..87C} .

In addition, we determine the oxygen abundance for each spaxel using the different calibrators presented in Sec.\ref{sec:calibrators}. The oxygen abundances presented in this article are determined as those at the effective radius $\rm{R_{eff}}$ from the best fitted radial gradient in each calibrator. Details about the method to derive abundance gradient, are described in \cite{2013A&A...554A..58S, 2016A&A...587A..70S}. The photometric properties of the galaxies (stellar mass [M$_{\odot}$], major position angle, and ellipticity) are obtained from the NSA catalog \footnote{\texttt{http://www.nsatlas.org}}. We use as our final sample those galaxies where it is possible to determine the oxygen abundance at the $\rm{R_{eff}}$ fulfilling the above criteria (i.e.,1704 objects). This selected sample corresponds mainly to galaxies located in the star-forming main sequence in the stellar mass - SFR plane (see Fig.\ref{fig:Sample}).

\subsection{Abundance Calibrators}
\label{sec:calibrators}

In order to avoid the controversy between different abundance calibrator derivations and to explore the MZR in the most general way we adopted a heterogeneous set of ten abundance calibrators. We derive the abundance using ({\it i}) calibrators based on the ''direct method", including the O3N2 and N2 line ratio calibrators proposed by \cite{2013A&A...559A.114M} (hereafter O3N2-M13 and N2, respectively); the calibration described in \cite{2010PhDT.......194R} (here after R23), and the calibrator proposed by \citep[][hereafter ONS]{2010ApJ...720.1738P}; ({\it ii}) an electronic-temperature corrected calibrator  proposed by \cite{2012ApJ...756L..14P} for an average of the abundances derived using the four previous methods (hereafter $t_2$); ({\it iii}) two mixed calibrators, based on the O3N2 calibrator \citep[][hereafter O3N2-PP04]{2004MNRAS.348L..59P}, and the R23 indicator \citep[][hereafter M08]{2008A&A...488..463M}; ({\it iv}) two calibrators based on pure photo-ionization models, the one included in the {\tt pyqz} code, which makes use of the O2, N2, S2, O3O2, O3N2, N2S2 and O3S2 line ratios as described in \citep[][hereafter pyqz]{2013ApJS..208...10D};  and the one adopted by \cite{2004ApJ...613..898T} in their exploration of the MZR based on the R23 line ratio (hereafter T04); and finally ({\it v}) a code to infer the metallicity from strong emission lines using Bayesian statistics  \citep[][hereafter, IZI]{2015ApJ...798...99B}.

\section{The MaNGA integrated MZ relation}
\label{sec:MZ}
\begin{figure*}
 \minipage{0.48\textwidth}
\includegraphics[width=\linewidth]{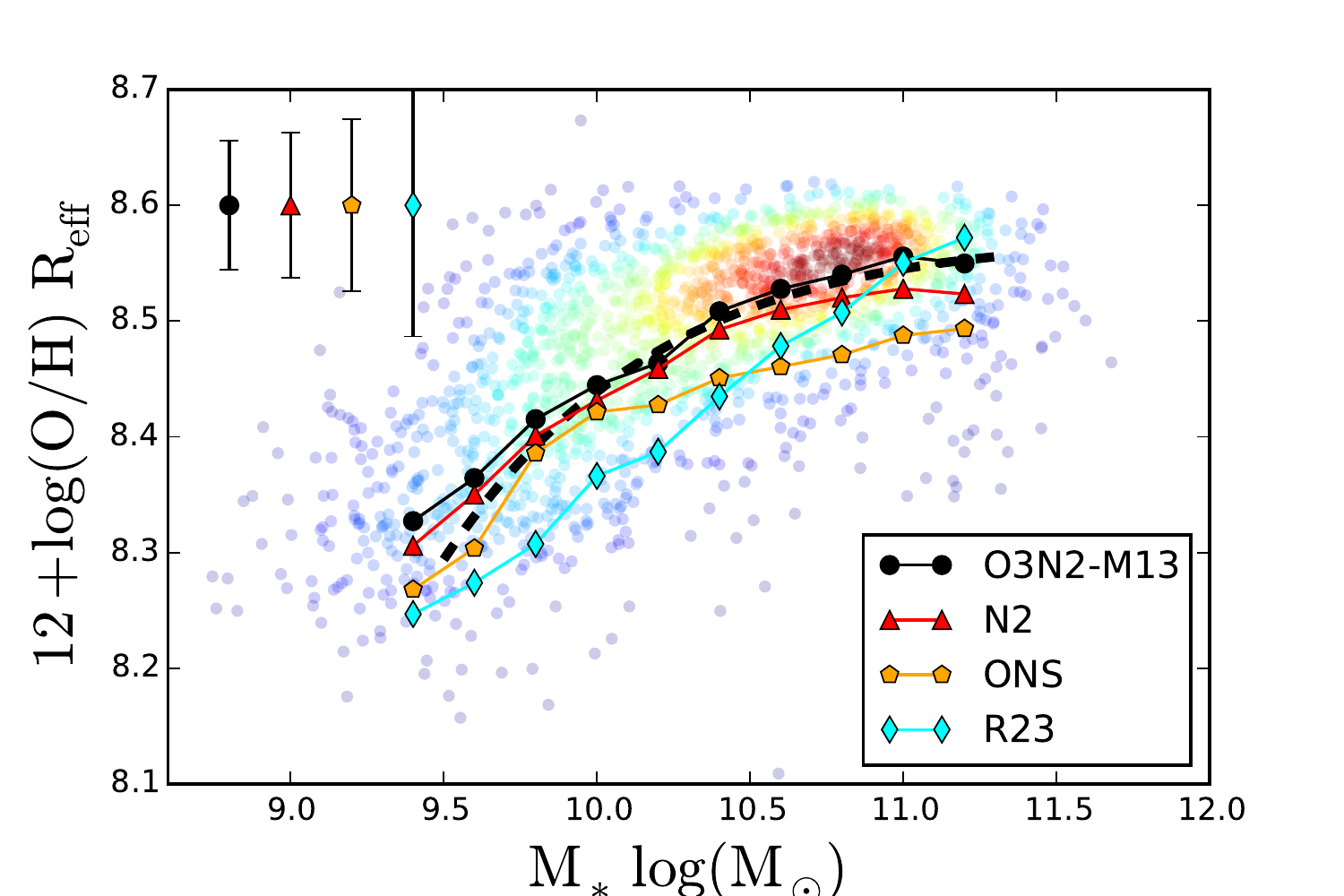}
 \endminipage
 \minipage{0.48\textwidth} 
\includegraphics[width=\linewidth]{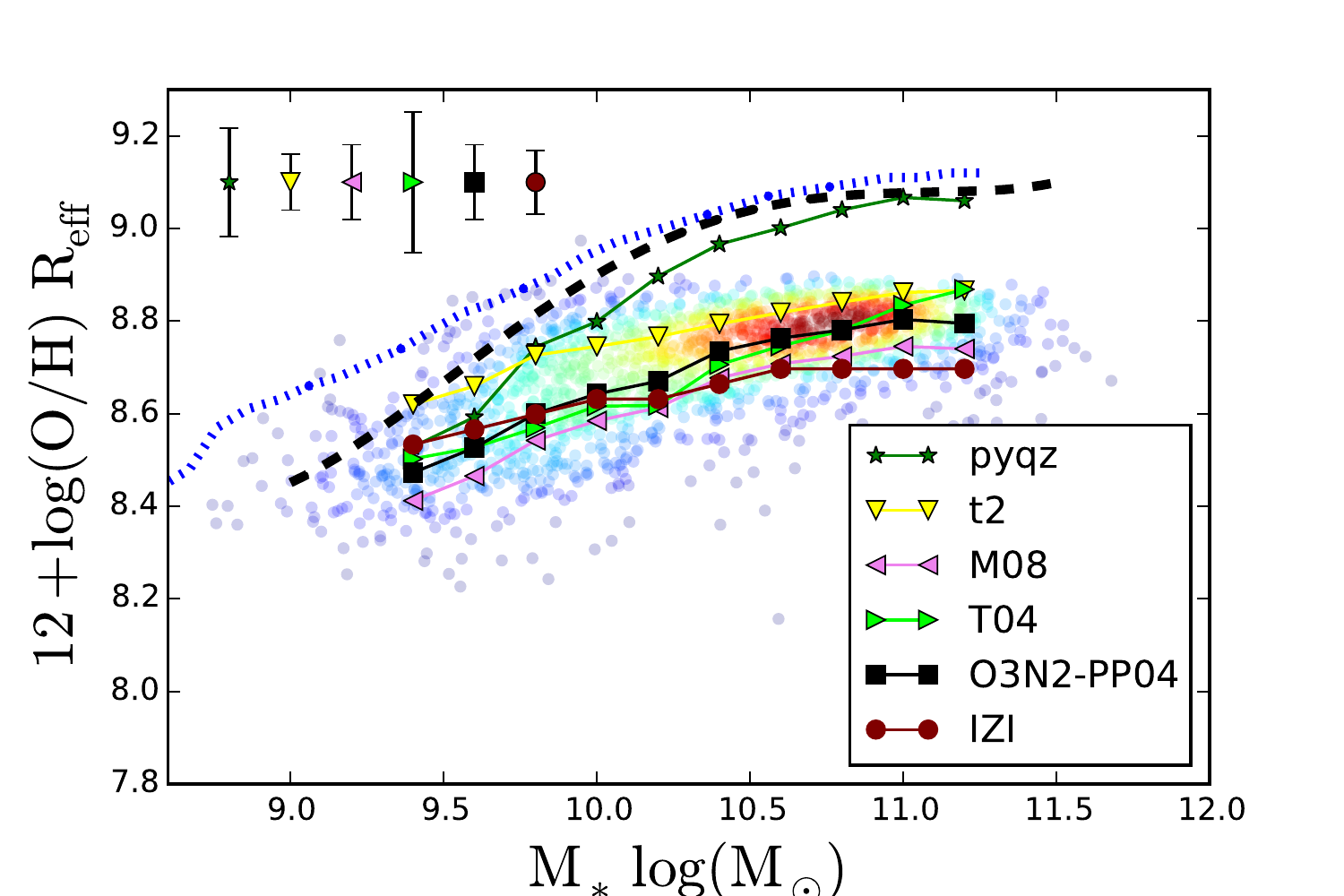}
\endminipage
\caption{MZR using different metallicity calibrators measured at $R_\mathrm{eff}$ for more than 1700 MaNGA galaxies. Line-connected symbols represent median values at given mass bin for the different calibrators. The error bars in the top-left symbols represent the average standard deviation for each indicator at different mass bins. In the left panel we use direct method-based metallicity calibrators (see Sec.~\ref{sec:calibrators} for their reference; O3N2-M13, black dots; N2, red triangles; ONS, orange pentagons; and R23, cyan diamonds). For reference of the dynamic range, we plot in the background the MZR for individual galaxies color-coded by their density using the O3N2-M13 calibrator. The dashed curve represents the best fit of the MZR  derived for the CALIFA sample \citep{2014A&A...563A..49S}. In the right panel we use semi-empirical or modeled calibrators (O3N2-P04, blue squares; {\tt pyqz}, green stars; {\it t2}, yellow down-triangles; M08, left-triangles; T04, right-triangles; IZI, maroon filled-circles). As for the left panel we plot the individual values for all the sample using O3N2-P04 indicator. The black dashed curve represents the best fit from \cite{2010MNRAS.408.2115M}. The dotted blue curve represents the median values from \cite{2004ApJ...613..898T}.} 
 \label{fig:MZ}
\end{figure*}

\begin{figure}
\includegraphics[width=\columnwidth]{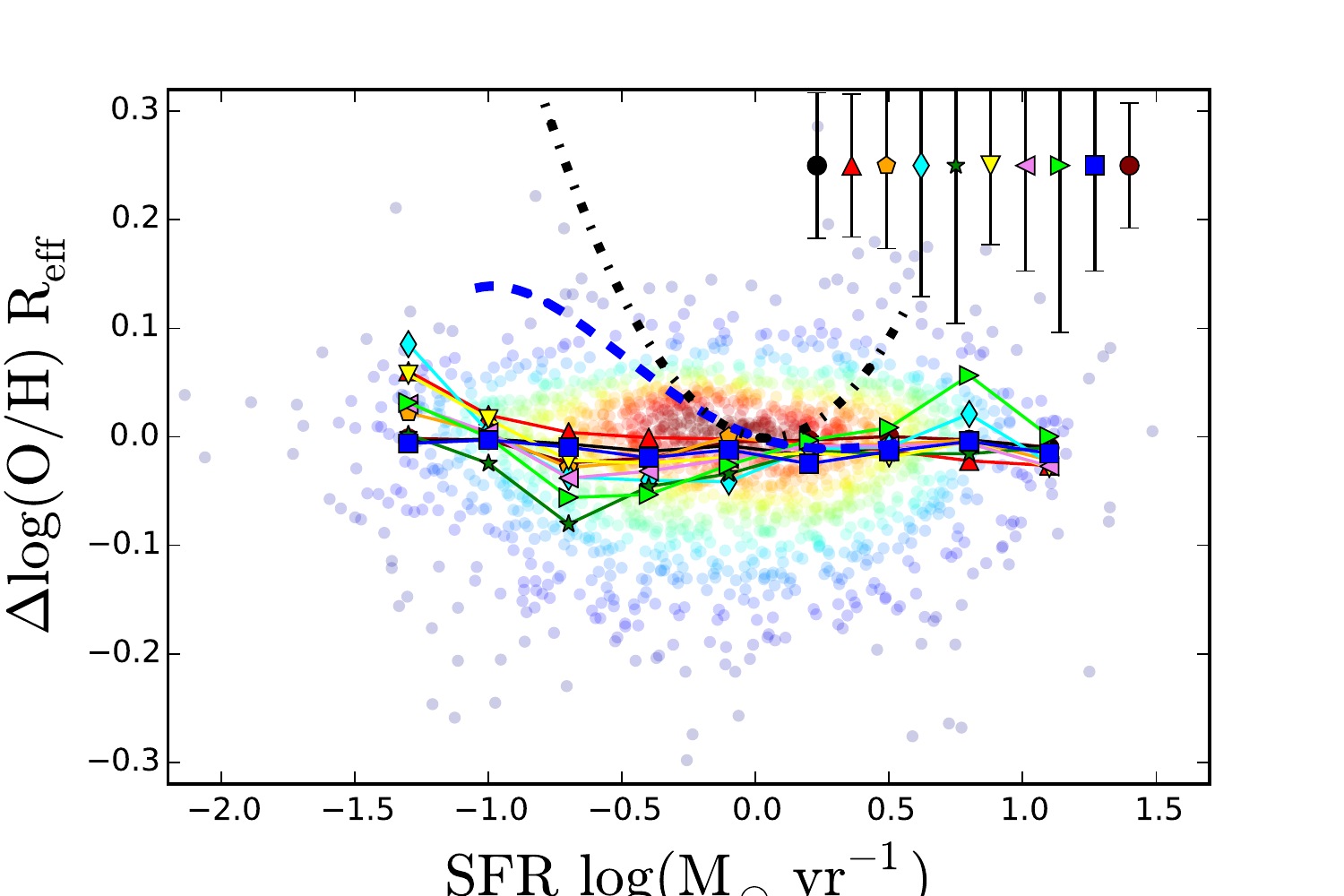}
\caption{MZR residuals from their best fitting curve against the SFR using different abundance indicators. Line-connected symbols represent median values at given SFR bin. The error bars in the top represent for each indicator the dispersion of the residuals from the best linear fit (see details in Sec.~\ref{sec:residuals}). For each calibrator  we use the same symbols as in Fig.\ref{fig:MZ}. For comparison we plot in the background the distribution of these residuals using the O3N2-M13 calibrator. The dashed line represents the relation between the scatter and the SFR presented in \cite{2010MNRAS.408.2115M} while the dotted line represent the relation by \cite{2010A&A...521L..53L}.} 
\label{fig:dMZ}
\end{figure}

\begin{figure}
\includegraphics[width=\columnwidth]{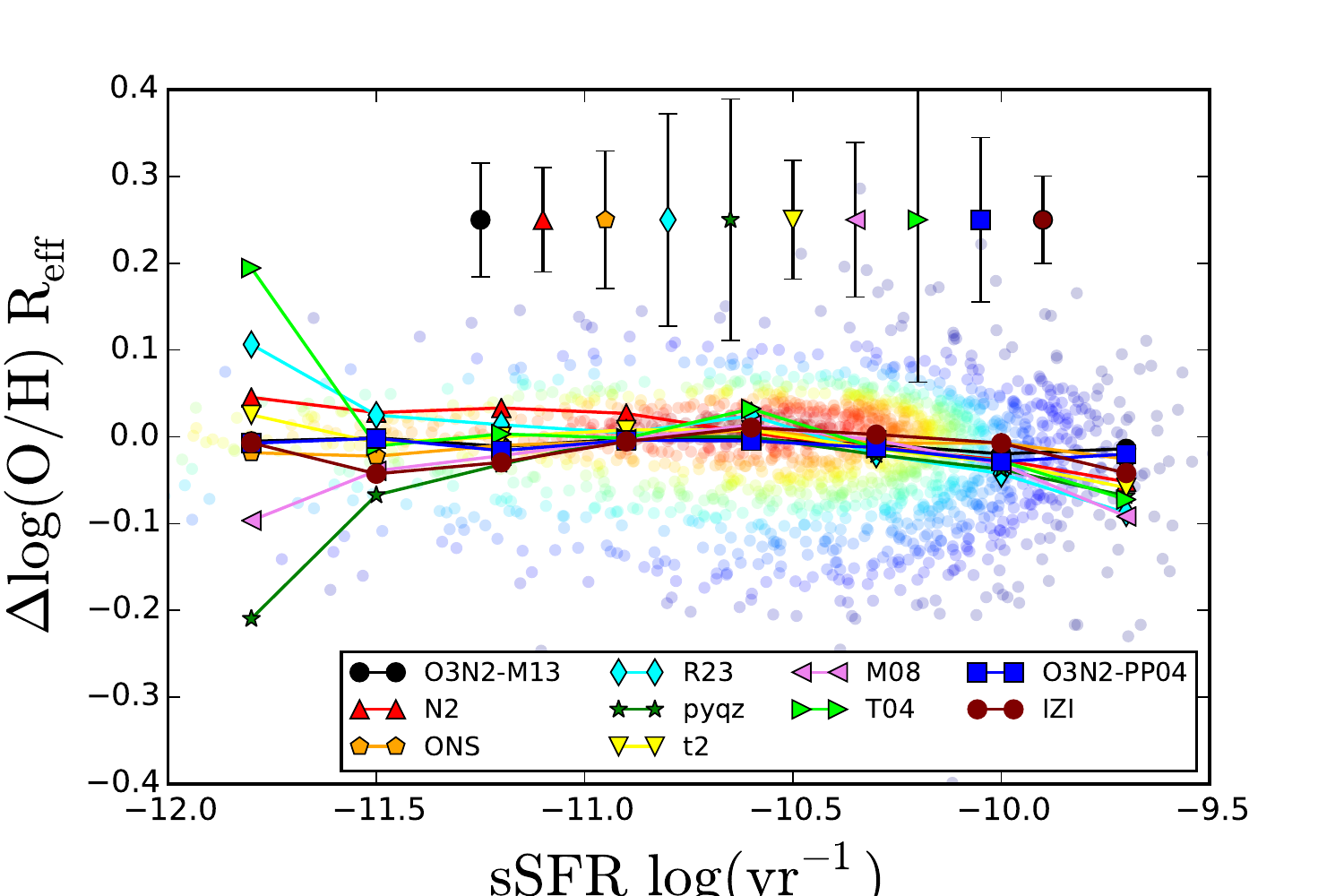}
\caption{MZR residuals from their best fitting curve as function of the sSFR. Line-connected symbols represent median values at given sSFR bin. The error bars in the bottom represent the dispersion of the residuals from the best linear fit. For each calibrator  we use the same symbols as in Fig.\ref{fig:MZ}.} 
\label{fig:dMZsSFR}
\end{figure}
%
\begin{figure*}
\includegraphics[width=\linewidth]{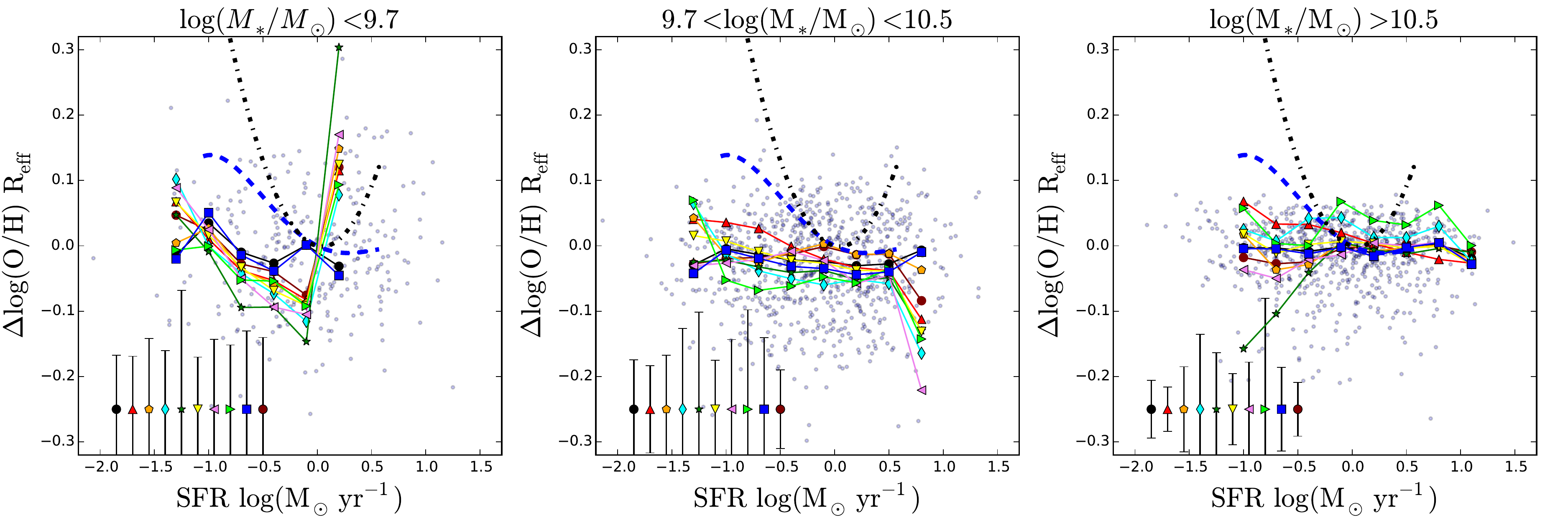}
\caption{MZR residuals from their best fitting curve as function of the SFR for different stellar mass bins. The sample has been binned in three stellar mass: low, intermediate and high stellar masses (left, middle, and right panels, respectively). Color code of the lines, symbols and error bars are similar as in Fig.\ref{fig:dMZ}. Data points represent the residuals from the O3N2-M13 calibrator.} 
\label{fig:dMZ_SFR_Mass}
\end{figure*}
\begin{figure*}
\includegraphics[width=\linewidth]{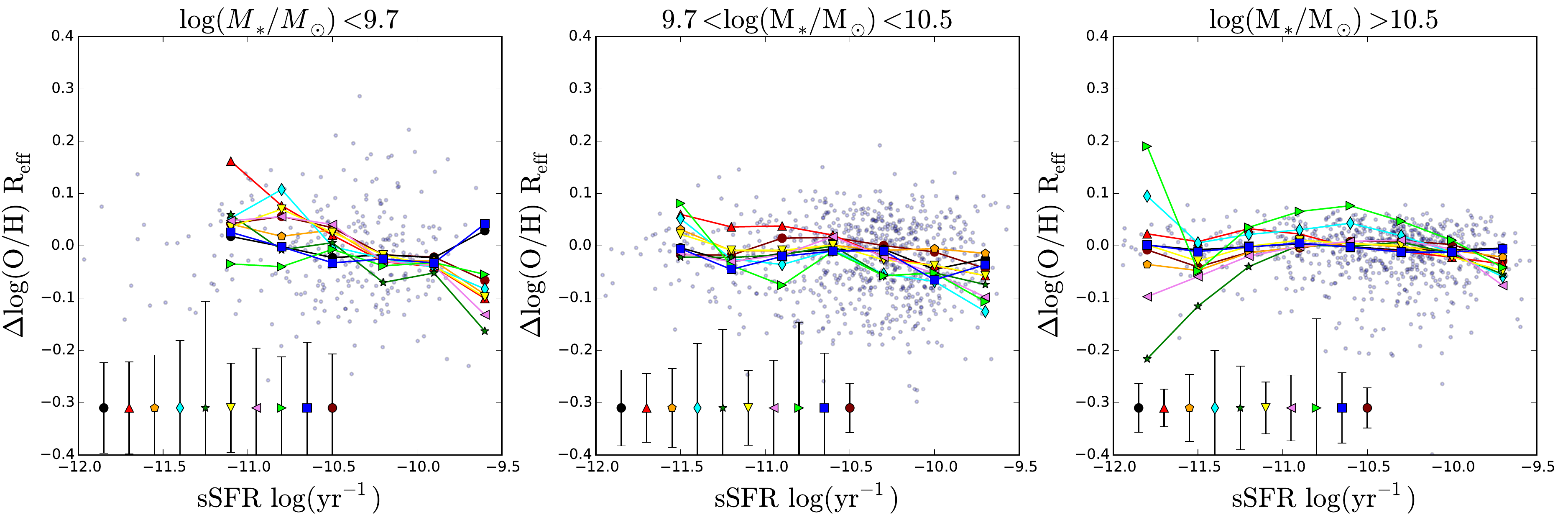}
\caption{MZR residuals from their best fitting curve as function of the sSFR for different stellar mass bins. The stellar mass bins are the same as in Fig.\ref{fig:dMZ_SFR_Mass}. Color code of the lines, symbols and error bars are similar as in Fig.\ref{fig:dMZsSFR}. Data points represent the residuals from the O3N2-M13 calibrator. } 
\label{fig:dMZ_sSFR_Mass}
\end{figure*}


In Fig.\ref{fig:MZ} we present the MZR for our sample of MaNGA targets using the ten different calibrators listed in the first column of Tab.~\ref{tab:val}. For visualization purposes, we plot the average metallicity at different stellar mass bins for our set of calibrators in two different panels. The left panel shows the MZR derived used only direct-method calibrators, whereas right panel shows mixed and photoionization model based calibrators. 

As we explain in Sect.\ref{sec:calibrators}, the derivation of these calibrators is quite heterogeneous. However, it is remarkable that metallicities derived from these different calibrators follow a similar trend. Metallicity increases with the stellar mass and reaching a constant value for more massive galaxies. Depending on the indicator, the absolute scale of the relation varies. 
For comparison, we  plot in left panel of Fig.~\ref{fig:MZ} the best-fit curve of the MZR derived from the CALIFA survey \citep[see dashed black curve,][]{2014A&A...563A..49S}. This curve is in excellent agreement with the MZR derived using direct method calibrators. We also plot in right panel of Fig.~\ref{fig:MZ} the best-fitted curve of the MZR derived by \cite{2010MNRAS.408.2115M} and the median values from \cite[][see their Table 3]{2004ApJ...613..898T} using single fiber spectroscopic data (see black-dashed and blue-dotted curves in left panel of Fig\ref{fig:MZ}, respectively). The curve derived by \cite{2010MNRAS.408.2115M} shows a very good agreement with the median points from the {\tt pyqz} calibrator (green-started points). On the other hand the curve from \cite{2004ApJ...613..898T} shows an offset to higher metallicities for low-mass galaxies.

\begin{table*}[!ht]
\footnotesize
\begin{center}
\begin{tabular} {c c c c c c c }
\toprule
Metallicity   & \multicolumn{2}{c}{MZ Best Fit} & $\sigma$ MZ-scatter & \multicolumn{2}{c}{$\Delta$MZ Best Fit } &  $\sigma \Delta$ MZ scatter  \\
 Indicator & $a$ (dex)  & $b$ (dex $/ \log(M_{\odot}))$ & (dex) & $\alpha$ (dex)  & $\beta$ (dex/$\log(M_{\odot} \, \mathrm{yr^{-1}}))$ & (dex) \\
\cmidrule{2-3}
\cmidrule{5-6}    
O3N2 - M13\,\, & 8.56 $\pm$ 0.02 & 0.005 $\pm$ 0.013 & 0.067 & -0.005 $\pm$ 0.1   &  0.001 $\pm$ 0.002 & 0.067 \\
N2             & 8.53 $\pm$ 0.01 & 0.01  $\pm$ 0.01  & 0.069 & -0.016 $\pm$ 0.02  & -0.013 $\pm$ 0.002 & 0.067 \\
ONS            & 8.49 $\pm$ 0.02 & 0.01  $\pm$ 0.02  & 0.075 & -0.027 $\pm$ 0.01  & -0.003 $\pm$ 0.001 & 0.075 \\
R23            & 8.53 $\pm$ 0.09 & 0.006 $\pm$ 0.052 & 0.125 & -0.053 $\pm$ 0.031 & -0.006 $\pm$ 0.003 & 0.125 \\ 
O3N2 P04       & 8.82 $\pm$ 0.03 & 0.007 $\pm$ 0.020 & 0.097 &  0.047 $\pm$ 0.039 &  0.007 $\pm$ 0.004 & 0.098 \\
pyqz           & 9.10 $\pm$ 0.05 & 0.013 $\pm$ 0.030 & 0.137 & -0.061 $\pm$ 0.021 & -0.006 $\pm$ 0.002 & 0.138 \\
t2             & 8.87 $\pm$ 0.03 & 0.005 $\pm$ 0.017 & 0.073 & -0.030 $\pm$ 0.048 & -0.003 $\pm$ 0.006 & 0.073 \\
M08            & 8.76 $\pm$ 0.03 & 0.007 $\pm$ 0.019 & 0.097 & -0.048 $\pm$ 0.018 & -0.005 $\pm$ 0.002 & 0.097 \\
T04            & 8.79 $\pm$ 0.13 & 0.006 $\pm$ 0.064 & 0.167 &  0.006 $\pm$ 0.006 &  0.001 $\pm$ 0.002 & 0.167 \\
IZI            & 8.71 $\pm$ 0.02 & 0.005 $\pm$ 0.019 & 0.059 &  0.008 $\pm$ 0.027 &  0.001 $\pm$ 0.003 & 0.059 \\
\bottomrule
\end{tabular}
\caption{Fitting parameters for the MZR and its scatter for the set of abundance calibrators used in this study. For each calibrator we list the: parameters $a$ and $b$ from the fitting of Eq.\ref{eq:fit} to the MZR;  $\sigma \Delta$ MZ scatter lists the standard deviation of the residuals after subtracting the best fit to the MZR; the parameters $\alpha$ and $\beta$ represent the linear fitting of the residuals of the MZR respect to the SFR (see Sec.\ref{sec:residuals}); $\sigma \Delta$ MZ scatter lists the standard deviation of the scatter of the linear fitting using the above parameters.}
\end{center}
\label{tab:val}
\end{table*}

\section{The SFR and sSFR dependence of the MZR residuals}
\label{sec:residuals}

In order to determine the residuals of the MZR for each indicator, we fit their median values at different mass bins using the functional form between these two parameters introduced by \cite{2011arXiv1112.3300M} and used by \cite{2013A&A...554A..58S}:
\begin{equation}
y=a+b(x-c)\exp(-(x-c))
\label{eq:fit}
\end{equation}
where $y = 12+\log(\mathrm{O/H})$ and $x=\log(M_*/ M_{\odot})-8.0$.  This functional form has been motivated by the shape of the MZR \citep{2013A&A...554A..58S}. The fitting coefficients, $a$, $b$ and $c$ represent the maximum metallicity, the curvature of the line and the stellar mass where the metallicity reach its maximum, respectively. We fix to $c = 4.0$ since at that mass the abundance is almost constant for any calibrator. In Tab.~\ref{tab:val} we present the best-fitted parameters $a$ and $b$ for all the calibrators. As expected from Fig.\ref{fig:MZ}, direct-method based calibrators show low values of $a$. The $b$ coefficient does not depend strongly on the calibrator, in fact within their uncertainties they are all similar.

We obtain the residuals of the MZR ($\Delta \log(\mathrm{O/H})$) for each calibrator by subtracting the metallicities from the best fitted curve. The standard deviation for each of them  ($\sigma$ MZ-scatter) is listed in  Tab.\ref{tab:val} (see also error bars in Fig.\ref{sec:MZ}).  We find that direct method, the $t2$ -based, and the {\tt IZI} Bayesian-based calibrators show lower dispersion in their scatter ($\sim$ 0.06 - 0.07 dex) compare to model-based or mixed calibrators ($\sim$ 0.10- 0.16 dex). This difference in the scatter of the MZR suggests that hybrid or model-based calibrators may introduce an artificial higher dynamical range of the residuals in comparison to the direct-methods and $t2$ calibrators. We also perform a similar analysis using as fitting function a fourth-order polynomial, following \cite{2010MNRAS.408.2115M}. We found similar standard deviations in the scatter as those reported in  Tab.\ref{tab:val} 

To explore the possible secondary dependence of the MZR with the SFR, it is necessary to study whether the residuals of the primary relation correlated with the SFR. In Fig.~\ref{fig:dMZ} we plot for each calibrator the median value of the $\Delta \log(\mathrm{O/H})$ within bins of SFR of 0.3 $\log(M_{\odot} \, \mathrm{yr^{-1}})$ width in a range of -1.5 and 1.0 $\log(M_{\odot} \, \mathrm{yr^{-1}})$. We find a good agreement of these medians for all the calibrators. Moreover, the dynamic range of these medians is smaller than $\pm$ 0.1 dex, which is the typical deviation of the MZR scatter. In other words, from these comparisons we do not find a significant trend of the scatter of the metallicity with respect to the SFR. Except for a very mild decrement of the residuals at low SFR, for some calibrators (e.g., N2, R23,and t2). Nevertheless, to quantify the possible linear relation of the scatter (i.e., a secondary relation of the SFR with respect to the MZR) we perform a linear fitting of these two parameters for our set of calibrators. In Tab.\ref{tab:val} we listed the best-fitted parameters ($\alpha$ and $\beta$ for the zero point and slope, respectively) for all the calibrators. We find that both the slope ($\alpha$) and the zero-point ($\beta$) are nearly zero. This is reinforcing evidence that there is not secondary relation of the MZR with the SFR. We also note that standard deviation of the residuals of this linear fitting is similar to the one derived from the MZR fitting (see last column in Tab.~\ref{tab:val}).

In Fig.~\ref{fig:dMZ} we compare two secondary relations reported in the literature with $\Delta \log(\mathrm{O/H})$ \citep[][blue dashed and black dotted lines, respectively]{2010MNRAS.408.2115M,2010A&A...521L..53L}. As prescribed by \cite{2010MNRAS.408.2115M}, we build the blue-dashed curve by subtracting their relation without SFR dependence  (i.e., $\mu_0$ in their Eq.4) to the same relation with SFR dependence (i.e., $\mu_{0.32}$). Similarly, we build the black-dotted curve by subtracting from the MZ-SFR relation with the one obtained removing the effect of the SFR in Eq. 1 in \cite{2010A&A...521L..53L}. The relation presented by \cite{2010MNRAS.408.2115M} highlights the fact that the secondary relation of the SFR is more evident at low stellar masses (see their Fig. 1) as observed in our plot (see blue-dashed line in Fig.\ref{fig:dMZ}). Although, it is evident that there is disagreement at low SFRs between this curve and our result, at larger SFRs this curve is a good representation of the scatter for almost all the calibrators. This in any case would happen if there is no dependence with the SFR at all. 
On the other hand, we do not find any correspondence between our data and the possible secondary relation of the SFR with the MZR described by the dotted black line in Fig.\ref{fig:dMZ} from \cite{2010A&A...521L..53L} at any SFR range, except at SFR~$\sim 1.0\,M_{\odot} \, \mathrm{yr^{-1}}$. In next section we explore if the presence of a relation of the SFR and $\Delta \log(\mathrm{O/H})$ can be observed at different stellar mass bins.

Finally, in Fig.~\ref{fig:dMZsSFR} we compare the residuals of the MZR with the sSFR. Except at the lowest bin of sSFR, we find that residuals do not vary significantly as function of the sSFR regardless of the abundance calibrator. This trend has also been observed in single-fiber spectroscopic studies. \cite{2004ApJ...613..898T} noted that the residuals of their derived MZR do not depend on the  EW(\Ha), which is a direct proxy for sSFR (see bottom-right panel in their Fig.7). For the lowest bin of sSFR the median  metallicities residuals change drastically depending on the calibrator. Some of them show  a positive residual (e.g., T04), zero (e.g., T04) or negative residuals (e.g., pyqz).
\subsection{Impact of stellar mass in the MZR residuals}
\label{dOH_mass}
The large sample of MaNGA galaxies allows us to investigate further the possible interplay of the stellar mass, and star formation in the metallicity. Even though we remove the stellar mass dependence in the metallicity by studying the residuals of the MZR, we still can ask how their observed relation (or lack thereof) with the SFR changes for different stellar masses. In Fig.~\ref{fig:dMZ_SFR_Mass} we plot $\Delta \log(\mathrm{O/H})$ against the SFR for three different mass bins from left to right: low, intermediate and high stellar mass galaxies. For the low-mass bin the residuals of the MZR seems to decrease with the SFR~$< 1.0\,M_{\odot} \, \mathrm{yr^{-1}}$. For larger SFRs the trend is not clear, some calibrators show larger residuals whereas the O3N2-based calibrators show residuals close to zero. In this low-mass bin for the range of SFRs where we observe the decrement in the residuals, we note that for a given SFR these residuals are smaller than the reported trend in \citep{2010MNRAS.408.2115M}. In fact, the slopes in log scale of the trend for all the calibrators are smaller than the slope presented in \citep{2010MNRAS.408.2115M} (0.03 to 0.23 dex / $M_{\odot}\,\mathrm{yr^{-1}}$). Even more, the standard deviation of the residuals distributions at low SFR bins is consistent with $\Delta \log(\mathrm{O/H}) = 0$ dex. The scarce fraction of galaxies with measured metallcities at high SFRs in this low-mass bin may explain the abrupt difference in the residuals at high SFR.

For the intermediate mass bin, the trends for the different calibrators are mixed. Some calibrators seem to decrease as the SFR increases (e.g., R23 and T04) while others show a constant trend around zero scatter (e.g., O3N2, ONS, and IZI). As we mentioned above most of the calibrators are not available for high SFR which may induce the strong decrement at high SFR. Without considering that last SFR bin, the distributions of all the calibrators are, within their deviations, consistent with a constant zero residual. As for the low-mass bin, the residual distributions at low SFRs is smaller than those reported in \citep{2010MNRAS.408.2115M}, except for the R23, pyqz, and T04 calibrators; where the deviation of their distributions are larger than the blue line.   

For the most massive galaxies (right panel in Fig.~\ref{fig:dMZ_SFR_Mass}) in most of the calibrator (except pyqz) the median residuals are rather constant around $\Delta \log(\mathrm{O/H}) \sim 0\,\mathrm{dex}$ for the different SFR bins. For the calibrators with the largest deviations (i.e., N2, ONS, and R23), the residual distributions include the trend described in blue dashed line, making statistically inconclusive whether or not the trend presented in \citep{2010MNRAS.408.2115M} can be described using these calibrators. 

In Fig.~\ref{fig:dMZ_sSFR_Mass} we study the residuals of the MZR with the sSFR at different stellar mass bins. For the low mass bin (left panel of Fig.~\ref{fig:dMZ_sSFR_Mass}), the observed trend of the  residuals with the sSFR depends strongly on the calibrator.  Some calibrators shows a significant decreasing in the residual with sSFR (e.g., N2), whereas some other do not exhibit an evident trend in the residuals with the SFR (e.g., O3N2-M13). A linear fitting of these two parameters  quantifies these trends from negative slopes to no trend (-0.17 to 0.001 dex $/ \log(\mathrm{yr^{-1}})$). As we note above for the relation of the MZR residuals and the SFR, the large distributions of the residuals at different bins of sSFR also make the observed negative trends compatible to the MZR residuals showing no trend with the sSFR. For the intermediate stellar masses (middle panel in Fig.~\ref{fig:dMZ_sSFR_Mass}) all the calibrators seem to show no relation between the residuals and the sSFR. For the most massive galaxies within the error bars the residuals show a lack of trend with the SFR. This further exploration of the MZR residual at different stellar masses suggests that our results in Fig.~\ref{fig:dMZ} and Fig.~\ref{fig:dMZsSFR} appear to be independent on the considered mass range. This is that metallicity does not strongly depends on the SFR.

\subsection{Central MaNGA Metallicities}
\label{sec:central}
\begin{figure}
\includegraphics[width=\linewidth]{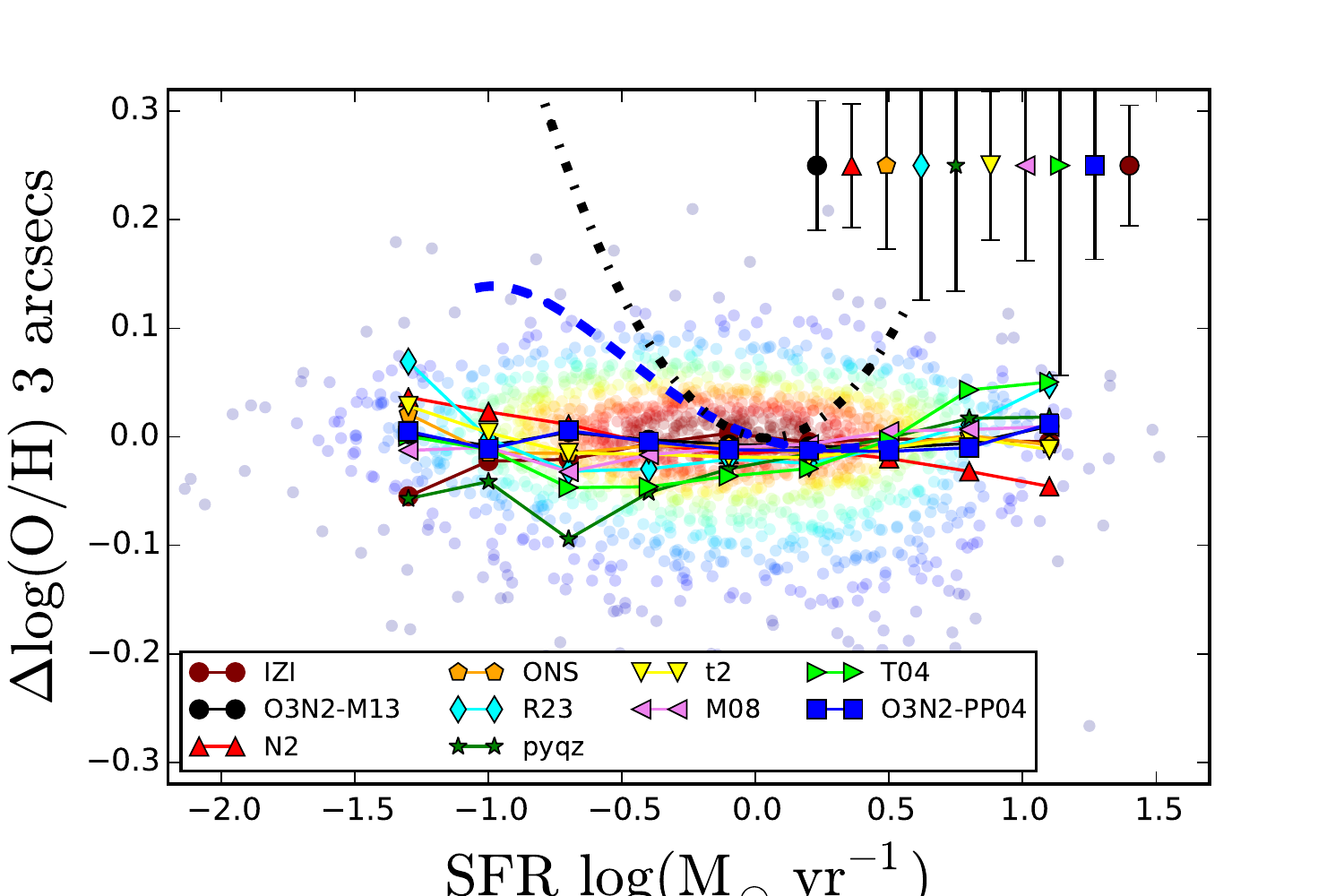}
\caption{Similar as Fig.~\ref{fig:dMZ}, with the derived MZR for all the metallicity calibrators and their corresponding residuals measured in a galactocentric aperture of 3 arcsec diameter.} 
\label{fig:dMZSFR_3arcsec}
\end{figure}
\begin{figure}
\includegraphics[width=\linewidth]{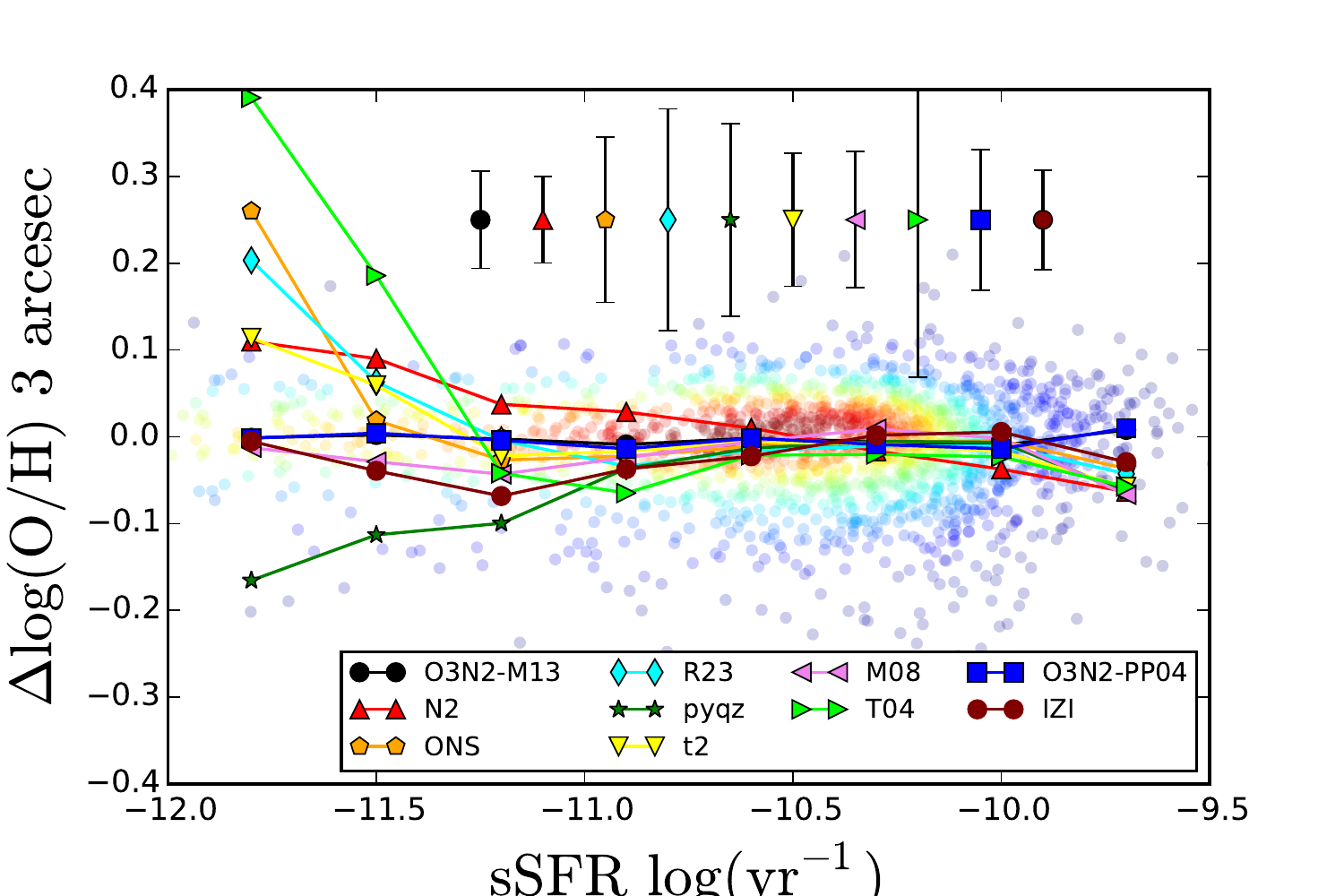}
\caption{Similar as Fig.~\ref{fig:dMZsSFR}, with the derived MZR for all the metallicity calibrators and their corresponding residuals measured in a galactocentric aperture of 3 arcsec diameter.} 
\label{fig:dMZsSFR_3arcsec}
\end{figure}

Spatially resolved observations allow us to measure physical properties at different galactic scales of our sample. Above, we analyze the metallicities derived at the effective radius using the fitting of the radial metallicity gradient. Nevertheless, we can select any aperture or region within the area covered by the field-of-view of the IFU instrument. One particular region of interest is the central 3-arcsec area of the observed galaxies. This is the same area cover by the single-fiber aperture from SDSS spectroscopic data. We perform the same analysis as in Sec.~\ref{sec:MZ} using the necessary fluxes to determine the metallicity from a 3 arcsec aperture centered in the optical nucleus of our sample of galaxies. Although not shown here, the MZR using the metallicity from this 3-arcsecs aperture is very similar to the one observed in Sec.~\ref{fig:MZ} for all the calibrators. Once we obtained the residuals of the MZR from the best fit using Eq.~\ref{eq:fit}, we proceed to compare these residuals with the SFR and sSFR following the same strategy as above. In Figs.~\ref{fig:dMZSFR_3arcsec}, and \ref{fig:dMZsSFR_3arcsec} we plot similar results  as in Figs.~\ref{fig:dMZ}, and  \ref{fig:dMZsSFR}, using instead the residuals of the MZR from the central 3 arcsec aperture. 

Comparing these residuals to those derived at the effective radius, we find that they show a slightly larger scatter among the different calibrators than the values derived at the effective radius when compared to the SFR (see Fig.~\ref{fig:dMZSFR_3arcsec}). This indicates that metallicity measurements at $\mathrm{R_{eff}}$ are better to characterize the entire galaxy's metallicity \citep[e.g.,][]{2014A&A...563A..49S,2017arXiv170309769S} . Similar as for the effective radius measurements, there are some calibrators that exhibit a very mild decrement of the MZR residuals as the SFR increases (e.g., N2, R23, and t2). In any case, the residuals from all the calibrators tend to have zero residuals at large SFR values. As for the metallicities measured at the effective radius, despite the mild decrement of these residuals at low SFR values, these median residuals are consistently lower than the values expected from the relations presented by \cite{2010MNRAS.408.2115M} and  \cite{2010A&A...521L..53L}.  

On the other hand, the trend from these same metallicity residuals  against the  sSFR differs considerably in comparison with those derived at $\mathrm{R_{eff}}$. Some calibrators show a clear trend with  residuals increasing as the sSFR decreases  ( T04, ONS, R23, and N2) where as  other remain  constant at different sSFR (O3N2-M13, O3N2-PP04, IZI, M08, and IZI). In summary, these results suggest that even when we consider the metallicity in a central 3-arcsecs aperture,  there seems to be no clear relation between the residuals and the SFR, even at low SFR regimes for any of the metallicity calibrator used in this study.  On the other hand, for the sSFR there seems to be some increment of the residuals at low sSFR at this particular aperture. However, as we seen above it depends strongly on the calibrator.

\section{Discussion}
\label{sec:Dis}

The main goal of this article is to shed some light in the question of whether or not there is a secondary relation in the already tight stellar mass - metallicity relation at global scales. With this idea in mind, we construct the MZR from a heterogenous set of ten metallcity calibrators. Thanks to the spatially resolved data provided by the MaNGA survey, all our metallicity measurements are derived at the effective radius of our sampled galaxies. In  Fig.\ref{fig:MZ} we present the MZR for the set of calibrators. As we noted in Sec.~\ref{sec:MZ}, apart from scaling factors, the trend observed for all the calibrators is similar; as the stellar mass increases the metallicity increases reaching a constant plateau. This remarkable result shows the robustness of the MZR regardless of the metallicity calibrator when measured at the effective radius. Using this result we explore the possible secondary relation of the (s)SFR with the metallicity residuals from the best fitted relation of the MZR for each calibrator.  As proposed by \citep{2010MNRAS.408.2115M} and \citep{2010A&A...521L..53L} if  there is a secondary relation of the (s)SFR with the MZR  one may expect a reduction in the scatter of the MZR when the SFR is introduced as a second parameter. In other words, if there is a secondary relation we should observe a clear trend between the residuals and the (s)SFR. Since the MZR shape appears to be generally independent on the calibrator, we also should expect that this possible trend between the above parameters to be independent on the calibrator. 

In Figs.~\ref{fig:dMZ} and \ref{fig:dMZsSFR}, we show the comparison of the MZR residuals against the SFR and sSFR for the set of ten calibrators, respetively. In Fig.~\ref{fig:dMZ} the discontinuous dotted and dashed lines represent two relations derived using single-fiber spectroscopy from the SDSS survey by \cite{2010A&A...521L..53L} and \cite{2010MNRAS.408.2115M}, respectively.  On the one hand, \cite{2010A&A...521L..53L} assumes that these three observables lay in a so called ``Fundamental Plane''. In other words, each of these observables can be described as a liner combination of the other two. These authors claim that for their observed parameters, this plane reduced the scatter from $\sim$ 0.26 as observed in their MZR to $\sim$ 0.16 dex as observed in the plane. The expected relation is very steep, decreasing more than 0.3 dex from SFR from 0.1 to 1 $\mathrm{M_{\odot}yr^{-1}}$. When we compare their expected metallicity residuals from this plane against the SFR with our observed residual, we are not able to reproduce them with any of the MZR residuals from our set of metallicity  calibrators.  Alternatively,  \cite{2010MNRAS.408.2115M} parametrized the metallicity as a function of a new parameter [$\mu_{\alpha} = \log(\mathrm{M_*/M_{\odot}}) + \alpha \log(\mathrm{SFR})$]. These authors shows that  $\alpha$  = -0.32 provides the best fit to their data.  This in turns suggest that in a given range of low-mass, low star-forming galaxies tend to have high metallicities.  Following this relation we should expect larger metallicity residuals in low star-forming galaxies than those with high SFRs, as represented by the blue dashed line in Fig.\ref{fig:dMZ}. For these authors, the MZR residuals decrease from $\sim$ 0.15 to 0.0 dex in a small SFR range (from 0.1 to 1 $\mathrm{M_{\odot}yr^{-1}}$). In contrast, we do not observed this trend with the same amplitude in the median MZR residuals from our set of calibrators. However, for some of these calibrators there seem to be a tendency of decreasing MZR residuals with the SFR (e.g., R23, N2, and t2).  
\begin{figure}
\includegraphics[width=\linewidth]{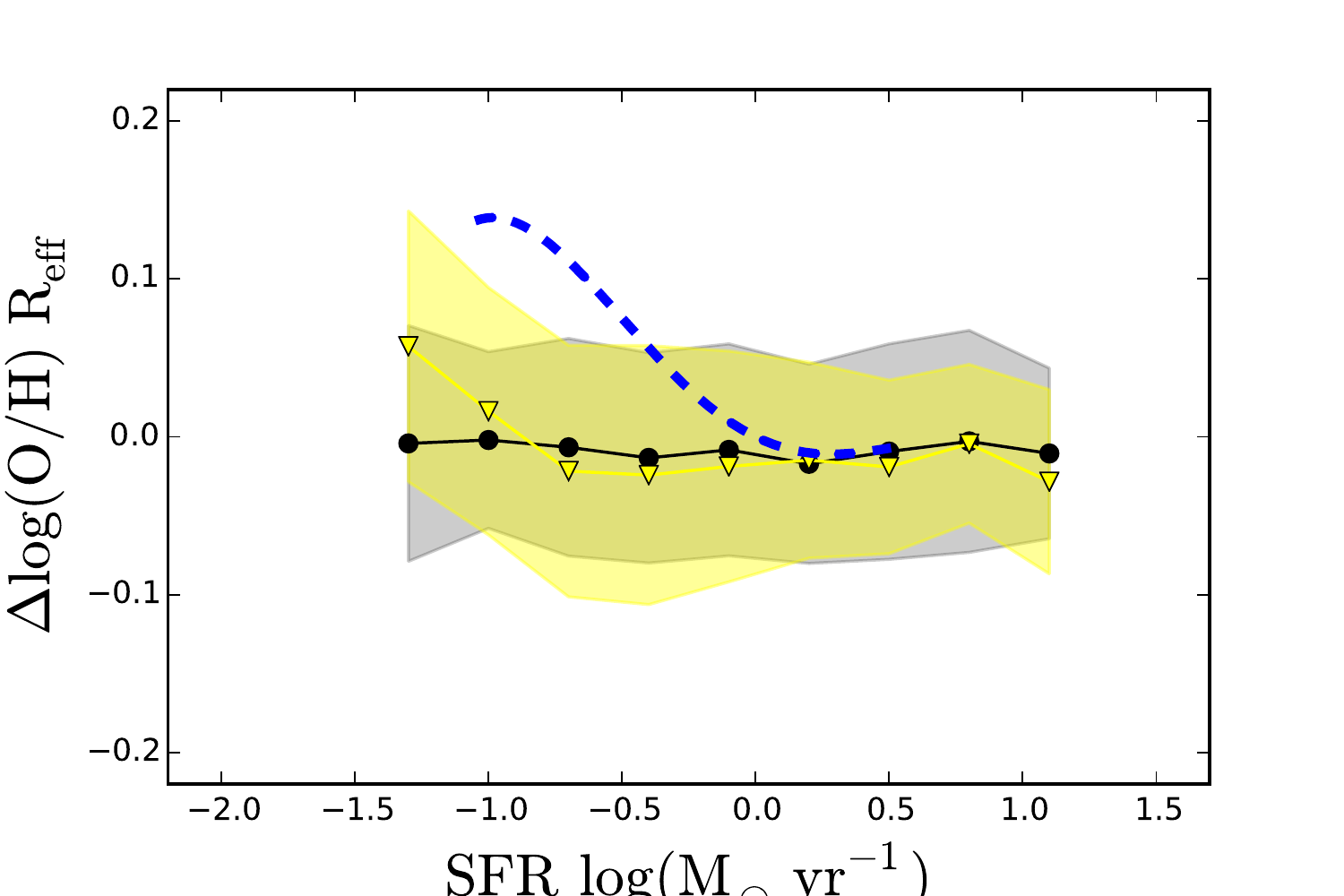}
\caption{MZR residuals against the SFR using two calibrators. Black and yellow points and lines represent the median residuals for the O3N2-M13 and t2 indicators, respectively. The shaded regions represent the area cover by the standard deviation. As in Fig.\ref{fig:dMZ}, the blue dotted line represent the \cite{2010MNRAS.408.2115M} relation.} 
\label{fig:comp}
\end{figure}
It may be the case that due to the statistical coverage, we are not able to sample enough the trend presented by \cite{2010MNRAS.408.2115M} at low SFR. In order to explore this, we compare in Fig.~\ref{fig:comp} the residuals of the calibrator with the tightest MZR relation (i.e., t2) along with one of the calibrators with almost zero MZR residual variations respect to the SFR (i.e., O3N2-M13). The yellow and black shades represent the standard deviation  of the residuals  cover by each of these calibrators at different SFR bins. For the t2 calibrator the distribution of metallicity residuals includes those expected by \cite{2010MNRAS.408.2115M}, although not for the entire SFR range. At the lowest SFR the expected relation from \cite{2010MNRAS.408.2115M} is not able to reproduce the t2 residual distributions. The same is true for the O3N2-M13, where the median residuals are almost constant. This comparision suggests that even if there is a secondary relation of the MZR with the SFR, it is weaker than those reported previously in the literature. It also indicates that this possible secondary relation may not be as robust as the MZR itself since it does depend on the selected metallicity calibrator. 

Except for the lowest sSFR bin, the MZR residuals are constant regardless of the metallicity calibrator (see, Fig.~\ref{fig:dMZsSFR}) . When we compare the residuals of the MZR with the sSFR we are comparing to quantities that in principle should not have any relation with the stellar mass, since both are normalized to this quantity. This in turn indicates that a secondary relation observed between these two quantities would be a strong indication of the actual impact of the SFR in the metal content of galaxies. Then, the result we observed in Fig.~\ref{fig:dMZsSFR} is also a very strong indication that metals does not seem to depend strongly on the SFR. 

In a recent article, \cite{2014ApJ...797..126S} found that the observed metallicity anti-correlates with the sSFR for SDSS galaxies located in the star-formation main sequence. To account for the impact of the stellar mass, they compare the metallicity and the sSFR for different mass bins. They stressed that the overall scatter from their observed Mass - Metallicity - SFR does not significantly reduced the scatter observed in their MZR. Our results agree qualitatively with the main conclusions from this study. We observe, at best, a weak dependence of the residuals of the metallicity (in other words, when we remove the stellar mass dependency on the metallicity) against the sSFR for the entire set of galaxies, regardless of the metallicity calibrator.  Once we divide our sample in stellar mass bins, we find trends of the residuals of the MZR with the SFR and sSFR (see Figs.~\ref{fig:dMZ_SFR_Mass} and \ref{fig:dMZ_sSFR_Mass}). However, these trends depend on the selected metallicity calibrator. A possible reason for the difference in these two studies is that we compare the residuals of the MZR instead of only the metallicity to the SFR. As we explain above, this ensures that we are comparing two quantities that are not heavily dependent on the stellar mass.  In conclusion, we suggest that previously observed trends between the metallicity and the sSFR could be induced by the distribution of galaxies into mass bins as well as the lack of proper subtraction of the dependence of the stellar mass in the metallicity.  

The spatially resolved data allows us to study in Sec.~\ref{sec:central}, the metallicity measured by the emission line fluxes integrated in a 3-arcsec aperture centered in the optical nucleus of our sample of galaxies. This emulates the metallicity data from a single fiber spectroscopy, such as the SDSS spectroscopic data. We perform the same analysis as at the effective radius (see Figs.~\ref{fig:dMZSFR_3arcsec} and ~\ref{fig:dMZsSFR_3arcsec}). The trend we observed between the MZR residuals and the SFR using this 3 arcsec aperture is quite similar to those observed at the effective radius. This suggests that the non-existent (or weak) relation of the MZR and the SFR is independent of the spatial scale. This result also is very robust for different metallicity calibrators.  Our data are also in agreement with recent results showing the lack or weak secondary relation of the MZR with the SFR for single-fiber aperture data. Using a single metallicity  calibrator, \cite{2016ApJ...823L..24K} showed that they were not able to reproduce the secondary relation between the SFR and the MZR proposed by \cite{2010MNRAS.408.2115M}. Ironically enough, as noted in their conclusions, the lack of a secondary relation was consistence with its existence. They claim that their metallicity calibrator is not sensitive to metallicity dilution or enhancement of the SFR due to metal-poor gas infall, therefore even if there was a secondary relation they were not  expecting not to observe it. On the other hand, \cite{2016ApJ...827...35T} explored systematic effects of the secondary relation between the MZR and the SFR. They found a weaker secondary relation in comparison to the one presented by \cite{2010MNRAS.408.2115M}. Along with these studies, our results show that even in the central region the assumed secondary relation with the SFR or the  sSFR, if exists, is much weaker than previously claimed.

The existence or not of an observational relation between these three parameters has a significant impact in how we understand the evolution and structure of galaxies in the Universe. Semi-analytical models as well as numerical simulations have explained this triple relation by invoking an interplay between global metal-poor gas inflow and outflows which remove enriched material far out the reach of the potential well of the galaxy \citep[e.g.,][, and references therein]{2016arXiv161200802F}. In a parallel work, we examine this possible secondary relation using the CALIFA spatially resolved dataset \citep{2017arXiv170309769S}. We find similar results to those presented in this study. Indeed, there is no clear trend nor statistical significant reduction of the scatter of the MZR by introducing a secondary dependance either with the SFR or the sSFR. The fact that we could not find a clear secondary relation either at the central region or at the effective radius indicates that the process responsable for the MZR seems to be scale independent.  We identify these results as the enrichment of the interstellar medium being dependent on local processes, with a strong dependence on the local star formation history, and local downsizing. Indeed, \cite{2012ApJ...756L..31R}  found an analogous local MZR between the surface mass density ($\Sigma_{*}$) and local metallicity. At low $\Sigma_{*}$ the local metallicity decreases, similar to the MZR for $\log(\mathrm{M_{*}/M_{\odot}}) <$ 10, indicating that metal enrichment and star formation is still occurring. At larger $\Sigma_{*}$ (or $\mathrm{M_{*}}$), both local and global metallicities reach a constant metallicity suggesting that in those regions (or galaxies)  star formation has ceased. Following this relation, it is also possible to reproduce the observed metallicity gradients in disk galaxies using the $\Sigma_{*}$ radial distributions \citep{2016MNRAS.463.2513B}. In conclusion, we suggest that global stellar mass - metallicity relation  is primarily  a consequence of the metal enrichment by the stellar population at local scales. Although we cannot completely rule out a dependency of the metallicity with the SFR, we find that if exists is much weaker than previously reported.

\section{Conclusions}
\label{sec:con}

We study the integrated stellar mass - metallicity relation (MZR) for more than 1700 galaxies included in the on-going SDSS-IV MaNGA survey. The wealth of this integral field spectroscopic data allows us to determine the metallicity at a fixed physical scale ($\mathrm{R_{eff}}$) using a set of ten calibrators covering a broad range of methods from empirical, mixed to pure photoionization models. We confirm for all calibrators the reported trend of the MZR using large samples of single-fiber spectroscopic data \citep[e.g.,][]{2004ApJ...613..898T, 2010MNRAS.408.2115M}, as well as those reported using integral-field spectroscopy \citep[e.g.,][]{2012ApJ...756L..31R, 2014A&A...563A..49S, 2016MNRAS.463.2513B} with different scaling factors \citep[e.g.,][]{2008ApJ...681.1183K}. Even more, we find that direct method and $t_2$ based calibrators show a MZR with smaller dispersion with respect to mixed and pure photoionization calibrators.  

We also explore a possible secondary relation of the MZR with the star formation rate (SFR) as well as with the specific SFR (sSFR). We find that residuals of the MZR do not show a evident correlation with SFR or with the sSFR regardless of the abundance calibrators used in this study. A further linear fitting of these residuals with the SFR or sSFR does not reduce the observed scatter. Even more, we note that the dispersion of these residuals is of the same order of magnitude as the errors of the adopted calibrators in most of the cases \citep[e.g.][]{2013A&A...559A.114M}, and therefore we do not expect any possible secondary relation in any case.  The above results are also valid when we consider the metallicity using fluxes integrated in a 3 arcesec aperture centered in their optical nuclei.

This is strong evidence supporting the lack of a secondary relation of the SFR with the MZR. Our results suggest that chemical enrichment in galaxies mainly occurs at local scales as proposed by \citet{2012ApJ...756L..31R, 2014A&A...563A..49S};  confirmed recently using data from the MaNGA survey  \citep{2016MNRAS.463.2513B}. This in turn also suggests that large-scale outflows do not appear to be the primary mechanism to enrich the ISM, whereas inflows seem to regulate the SFR \citep{2013ApJ...772..119L}. 

Finally, we note that a local/global SFR - stellar mass sequence \citep[SFMS, e.g.,][]{2013A&A...554A..58S,2016arXiv160202770C} implies that a possible secondary relation of the MZR with the SFR would be just a re-scaling of the stellar mass-axis, rather than a reduction in the scatter of the MZR.

\section*{Acknowledgements}
SFS thanks the ConaCyt programs IA-180125 and DGAPA IA100815
and IA101217 for their support to this project. G.B. is supported by CONICYT/FONDECYT, Programa de Iniciacion, Folio 11150220. Funding for the Sloan Digital Sky Survey IV has been provided by
the Alfred P. Sloan Foundation, the U.S. Department of Energy Office of
Science, and the Participating Institutions. SDSS-IV acknowledges
support and resources from the Center for High-Performance Computing at
the University of Utah. The SDSS web site is www.sdss.org.

SDSS-IV is managed by the Astrophysical Research Consortium for the 
Participating Institutions of the SDSS Collaboration including the 
Brazilian Participation Group, the Carnegie Institution for Science, 
Carnegie Mellon University, the Chilean Participation Group, the French Participation Group, Harvard-Smithsonian Center for Astrophysics, 
Instituto de Astrof\'isica de Canarias, The Johns Hopkins University, 
Kavli Institute for the Physics and Mathematics of the Universe (IPMU) / 
University of Tokyo, Lawrence Berkeley National Laboratory, 
Leibniz Institut f\"ur Astrophysik Potsdam (AIP),  
Max-Planck-Institut f\"ur Astronomie (MPIA Heidelberg), 
Max-Planck-Institut f\"ur Astrophysik (MPA Garching), 
Max-Planck-Institut f\"ur Extraterrestrische Physik (MPE), 
National Astronomical Observatories of China, New Mexico State University, 
New York University, University of Notre Dame, 
Observat\'ario Nacional / MCTI, The Ohio State University, 
Pennsylvania State University, Shanghai Astronomical Observatory, 
United Kingdom Participation Group,
Universidad Nacional Aut\'onoma de M\'exico, University of Arizona, 
University of Colorado Boulder, University of Oxford, University of Portsmouth, 
University of Utah, University of Virginia, University of Washington, University of Wisconsin, 
Vanderbilt University, and Yale University.

\bibliographystyle{mnras}
\bibliography{main}
\end{document}